\documentclass[10pt, conference]{IEEEtran}
\IEEEoverridecommandlockouts
\usepackage{cite}
\usepackage{amsmath,amssymb,amsfonts}
\usepackage{algorithmic}
\usepackage{graphicx}
\usepackage{textcomp}
\usepackage{xcolor}
\usepackage{tikz}
\usepackage{colortbl}
\usepackage{booktabs}
\def\BibTeX{{\rm B\kern-.05em{\sc i\kern-.025em b}\kern-.08em
    T\kern-.1667em\lower.7ex\hbox{E}\kern-.125emX}}

\usepackage[left=1.7cm,right=1.7cm,top=1.9cm]{geometry}

\usepackage{soul}

\usepackage{glossaries}
\setacronymstyle{long-short}
\newacronym{3gpp}{3GPP}{3rd Generation Partnership Project}
\newacronym{4g}{4G}{4th generation}
\newacronym{5g}{5G}{5th generation}
\newacronym{6g}{6G}{6th generation}
\newacronym{5gc}{5GC}{5G Core}
\newacronym{adc}{ADC}{Analog to Digital Converter}
\newacronym{aerpaw}{AERPAW}{Aerial Experimentation and Research Platform for Advanced Wireless}
\newacronym{ai}{AI}{Artificial Intelligence}
\newacronym{aimd}{AIMD}{Additive Increase Multiplicative Decrease}
\newacronym{am}{AM}{Acknowledged Mode}
\newacronym{amc}{AMC}{Adaptive Modulation and Coding}
\newacronym{amf}{AMF}{Access and Mobility Management Function}
\newacronym{aops}{AOPS}{Adaptive Order Prediction Scheduling}
\newacronym{api}{API}{Application Programming Interface}
\newacronym{apn}{APN}{Access Point Name}
\newacronym{ap}{AP}{Application Protocol}
\newacronym{aqm}{AQM}{Active Queue Management}
\newacronym{ausf}{AUSF}{Authentication Server Function}
\newacronym{avc}{AVC}{Advanced Video Coding}
\newacronym{awgn}{AGWN}{Additive White Gaussian Noise}
\newacronym{balia}{BALIA}{Balanced Link Adaptation Algorithm}
\newacronym{bbu}{BBU}{Base Band Unit}
\newacronym{bdp}{BDP}{Bandwidth-Delay Product}
\newacronym{ber}{BER}{Bit Error Rate}
\newacronym{bf}{BF}{Beamforming}
\newacronym{bler}{BLER}{Block Error Rate}
\newacronym{brr}{BRR}{Bayesian Ridge Regressor}
\newacronym{bs}{BS}{Base Station}
\newacronym{bsr}{BSR}{Buffer Status Report}
\newacronym{bss}{BSS}{Business Support System}
\newacronym{ca}{CA}{Carrier Aggregation}
\newacronym{caas}{CaaS}{Connectivity-as-a-Service}
\newacronym{cb}{CB}{Code Block}
\newacronym{cc}{CC}{Congestion Control}
\newacronym{ccid}{CCID}{Congestion Control ID}
\newacronym{cco}{CC}{Carrier Component}
\newacronym{cdd}{CDD}{Cyclic Delay Diversity}
\newacronym{cdf}{CDF}{Cumulative Distribution Function}
\newacronym{cdn}{CDN}{Content Distribution Network}
\newacronym{cli}{CLI}{Command-line Interface}
\newacronym{cn}{CN}{Core Network}
\newacronym{codel}{CoDel}{Controlled Delay Management}
\newacronym{comac}{COMAC}{Converged Multi-Access and Core}
\newacronym{cord}{CORD}{Central Office Re-architected as a Datacenter}
\newacronym{cornet}{CORNET}{COgnitive Radio NETwork}
\newacronym{cosmos}{COSMOS}{Cloud Enhanced Open Software Defined Mobile Wireless Testbed for City-Scale Deployment}
\newacronym{cots}{COTS}{Commercial Off-the-Shelf}
\newacronym{cp}{CP}{Control Plane}
\newacronym{cyp}{CP}{Cyclic Prefix}
\newacronym{up}{UP}{User Plane}
\newacronym{cpu}{CPU}{Central Processing Unit}
\newacronym{cqi}{CQI}{Channel Quality Information}
\newacronym{cr}{CR}{Cognitive Radio}
\newacronym{cran}{CRAN}{Cloud \gls{ran}}
\newacronym{crs}{CRS}{Cell Reference Signal}
\newacronym{csi}{CSI}{Channel State Information}
\newacronym{csirs}{CSI-RS}{Channel State Information - Reference Signal}
\newacronym{cu}{CU}{Central Unit}
\newacronym{d2tcp}{D$^2$TCP}{Deadline-aware Data center TCP}
\newacronym{d3}{D$^3$}{Deadline-Driven Delivery}
\newacronym{dac}{DAC}{Digital to Analog Converter}
\newacronym{dag}{DAG}{Directed Acyclic Graph}
\newacronym{das}{DAS}{Distributed Antenna System}
\newacronym{dash}{DASH}{Dynamic Adaptive Streaming over HTTP}
\newacronym{dc}{DC}{Dual Connectivity}
\newacronym{dccp}{DCCP}{Datagram Congestion Control Protocol}
\newacronym{dce}{DCE}{Direct Code Execution}
\newacronym{dci}{DCI}{Downlink Control Information}
\newacronym{dctcp}{DCTCP}{Data Center TCP}
\newacronym{dl}{DL}{Downlink}
\newacronym{dmr}{DMR}{Deadline Miss Ratio}
\newacronym{dmrs}{DMRS}{DeModulation Reference Signal}
\newacronym{drlcc}{DRL-CC}{Deep Reinforcement Learning Congestion Control}
\newacronym{drs}{DRS}{Discovery Reference Signal}
\newacronym{du}{DU}{Distributed Unit}
\newacronym{e2e}{E2E}{end-to-end}
\newacronym{earfcn}{EARFCN}{E-UTRA Absolute Radio Frequency Channel Number}
\newacronym{ecaas}{ECaaS}{Edge-Cloud-as-a-Service}
\newacronym{ecn}{ECN}{Explicit Congestion Notification}
\newacronym{edf}{EDF}{Earliest Deadline First}
\newacronym{embb}{eMBB}{Enhanced Mobile Broadband}
\newacronym{empower}{EMPOWER}{EMpowering transatlantic PlatfOrms for advanced WirEless Research}
\newacronym{enb}{eNB}{evolved Node Base}
\newacronym{endc}{EN-DC}{E-UTRAN-\gls{nr} \gls{dc}}
\newacronym{epc}{EPC}{Evolved Packet Core}
\newacronym{eps}{EPS}{Evolved Packet System}
\newacronym{es}{ES}{Edge Server}
\newacronym{etsi}{ETSI}{European Telecommunications Standards Institute}
\newacronym[firstplural=Estimated Times of Arrival (ETAs)]{eta}{ETA}{Estimated Time of Arrival}
\newacronym{eutran}{E-UTRAN}{Evolved Universal Terrestrial Access Network}
\newacronym{faas}{FaaS}{Function-as-a-Service}
\newacronym{fapi}{FAPI}{Functional Application Platform Interface}
\newacronym{fdd}{FDD}{Frequency Division Duplexing}
\newacronym{fdm}{FDM}{Frequency Division Multiplexing}
\newacronym{fdma}{FDMA}{Frequency Division Multiple Access}
\newacronym{fed4fire}{FED4FIRE+}{Federation 4 Future Internet Research and Experimentation Plus}
\newacronym{fir}{FIR}{Finite Impulse Response}
\newacronym{fit}{FIT}{Future \acrlong{iot}}
\newacronym{fpga}{FPGA}{Field Programmable Gate Array}
\newacronym{fr2}{FR2}{Frequency Range 2}
\newacronym{fs}{FS}{Fast Switching}
\newacronym{fscc}{FSCC}{Flow Sharing Congestion Control}
\newacronym{ftp}{FTP}{File Transfer Protocol}
\newacronym{fw}{FW}{Flow Window}
\newacronym{ge}{GE}{Gaussian Elimination}
\newacronym{gnb}{gNB}{Next Generation Node Base}
\newacronym{gop}{GOP}{Group of Pictures}
\newacronym{gpr}{GPR}{Gaussian Process Regressor}
\newacronym{gpu}{GPU}{Graphics Processing Unit}
\newacronym{gtp}{GTP}{GPRS Tunneling Protocol}
\newacronym{gtpc}{GTP-C}{GPRS Tunnelling Protocol Control Plane}
\newacronym{gtpu}{GTP-U}{GPRS Tunnelling Protocol User Plane}
\newacronym{gtpv2c}{GTPv2-C}{\gls{gtp} v2 - Control}
\newacronym{gw}{GW}{Gateway}
\newacronym{harq}{HARQ}{Hybrid Automatic Repeat reQuest}
\newacronym{hetnet}{HetNet}{Heterogeneous Network}
\newacronym{hh}{HH}{Hard Handover}
\newacronym{hol}{HOL}{Head-of-Line}
\newacronym{hqf}{HQF}{Highest-quality-first}
\newacronym{hss}{HSS}{Home Subscription Server}
\newacronym{http}{HTTP}{HyperText Transfer Protocol}
\newacronym{ia}{IA}{Initial Access}
\newacronym{iab}{IAB}{Integrated Access and Backhaul}
\newacronym{ic}{IC}{Incident Command}
\newacronym{ietf}{IETF}{Internet Engineering Task Force}
\newacronym{imsi}{IMSI}{International Mobile Subscriber Identity}
\newacronym{imt}{IMT}{International Mobile Telecommunication}
\newacronym{iot}{IoT}{Internet of Things}
\newacronym{ip}{IP}{Internet Protocol}
\newacronym{itu}{ITU}{International Telecommunication Union}
\newacronym{kpi}{KPI}{Key Performance Indicator}
\newacronym{kpm}{KPM}{Key Performance Measurement}
\newacronym{kvm}{KVM}{Kernel-based Virtual Machine}
\newacronym{los}{LOS}{Line-of-Sight}
\newacronym{lsm}{LSM}{Link-to-System Mapping}
\newacronym{lstm}{LSTM}{Long Short Term Memory}
\newacronym{lte}{LTE}{Long Term Evolution}
\newacronym{lxc}{LXC}{Linux Container}
\newacronym{m2m}{M2M}{Machine to Machine}
\newacronym{mac}{MAC}{Medium Access Control}
\newacronym{manet}{MANET}{Mobile Ad Hoc Network}
\newacronym{mano}{MANO}{Management and Orchestration}
\newacronym{mc}{MC}{Multi-Connectivity}
\newacronym{mcc}{MCC}{Mobile Cloud Computing}
\newacronym{mchem}{MCHEM}{Massive Channel Emulator}
\newacronym{mcs}{MCS}{Modulation and Coding Scheme}
\newacronym{mec}{MEC}{Multi-access Edge Computing}
\newacronym{mec2}{MEC}{Mobile Edge Cloud}
\newacronym{mfc}{MFC}{Mobile Fog Computing}
\newacronym{mgen}{MGEN}{Multi-Generator}
\newacronym{mi}{MI}{Mutual Information}
\newacronym{mib}{MIB}{Master Information Block}
\newacronym{miesm}{MIESM}{Mutual Information Based Effective SINR}
\newacronym{mimo}{MIMO}{Multiple Input, Multiple Output}
\newacronym{ml}{ML}{Machine Learning}
\newacronym{mlr}{MLR}{Maximum-local-rate}
\newacronym[plural=\gls{mme}s,firstplural=Mobility Management Entities (MMEs)]{mme}{MME}{Mobility Management Entity}
\newacronym{mmtc}{mMTC}{Massive Machine-Type Communications}
\newacronym{mmwave}{mmWave}{millimeter wave}
\newacronym{mpdccp}{MP-DCCP}{Multipath Datagram Congestion Control Protocol}
\newacronym{mptcp}{MPTCP}{Multipath TCP}
\newacronym{mr}{MR}{Maximum Rate}
\newacronym{mrdc}{MR-DC}{Multi \gls{rat} \gls{dc}}
\newacronym{mse}{MSE}{Mean Square Error}
\newacronym{mss}{MSS}{Maximum Segment Size}
\newacronym{mt}{MT}{Mobile Termination}
\newacronym{mtd}{MTD}{Machine-Type Device}
\newacronym{mtu}{MTU}{Maximum Transmission Unit}
\newacronym{mumimo}{MU-MIMO}{Multi-user \gls{mimo}}
\newacronym{mvno}{MVNO}{Mobile Virtual Network Operator}
\newacronym{nalu}{NALU}{Network Abstraction Layer Unit}
\newacronym{nas}{NAS}{Network Attached Storage}
\newacronym{nat}{NAT}{Network Address Translation}
\newacronym{nbiot}{NB-IoT}{Narrow Band IoT}
\newacronym{nfv}{NFV}{Network Function Virtualization}
\newacronym{nfvi}{NFVI}{Network Function Virtualization Infrastructure}
\newacronym{ni}{NI}{Network Interfaces}
\newacronym{nic}{NIC}{Network Interface Card}
\newacronym{nlos}{NLOS}{Non-Line-of-Sight}
\newacronym{now}{NOW}{Non Overlapping Window}
\newacronym{nsm}{NSM}{Network Service Mesh}
\newacronym{nrf}{NRF}{Network Repository Function}
\newacronym{nsa}{NSA}{Non Stand Alone}
\newacronym{nse}{NSE}{Network Slicing Engine}
\newacronym{nssf}{NSSF}{Network Slice Selection Function}
\newacronym{o2i}{O2I}{Outdoor to Indoor}
\newacronym{oai}{OAI}{OpenAirInterface}
\newacronym{oaicn}{OAI-CN}{\gls{oai} \acrlong{cn}}
\newacronym{oairan}{OAI-RAN}{\acrlong{oai} \acrlong{ran}}
\newacronym{oam}{OAM}{Operations, Administration and Maintenance}
\newacronym{ofdm}{OFDM}{Orthogonal Frequency Division Multiplexing}
\newacronym{olia}{OLIA}{Opportunistic Linked Increase Algorithm}
\newacronym{omec}{OMEC}{Open Mobile Evolved Core}
\newacronym{onap}{ONAP}{Open Network Automation Platform}
\newacronym{onf}{ONF}{Open Networking Foundation}
\newacronym{onos}{ONOS}{Open Networking Operating System}
\newacronym{oom}{OOM}{\gls{onap} Operations Manager}
\newacronym{opnfv}{OPNFV}{Open Platform for \gls{nfv}}
\newacronym{orbit}{ORBIT}{Open-Access Research Testbed for Next-Generation Wireless Networks}
\newacronym{os}{OS}{Operating System}
\newacronym{oss}{OSS}{Operations Support System}
\newacronym{pa}{PA}{Position-aware}
\newacronym{pase}{PASE}{Prioritization, Arbitration, and Self-adjusting Endpoints}
\newacronym{pawr}{PAWR}{Platforms for Advanced Wireless Research}
\newacronym{pbch}{PBCH}{Physical Broadcast Channel}
\newacronym{pcef}{PCEF}{Policy and Charging Enforcement Function}
\newacronym{pcfich}{PCFICH}{Physical Control Format Indicator Channel}
\newacronym{pcrf}{PCRF}{Policy and Charging Rules Function}
\newacronym{pdcch}{PDCCH}{Physical Downlink Control Channel}
\newacronym{pdcp}{PDCP}{Packet Data Convergence Protocol}
\newacronym{pdsch}{PDSCH}{Physical Downlink Shared Channel}
\newacronym{pdu}{PDU}{Packet Data Unit}
\newacronym{pf}{PF}{Proportional Fair}
\newacronym{pgw}{PGW}{Packet Gateway}
\newacronym{phich}{PHICH}{Physical Hybrid ARQ Indicator Channel}
\newacronym{phy}{PHY}{Physical}
\newacronym{pmch}{PMCH}{Physical Multicast Channel}
\newacronym{pmi}{PMI}{Precoding Matrix Indicators}
\newacronym{powder}{POWDER}{Platform for Open Wireless Data-driven Experimental Research}
\newacronym{ppo}{PPO}{Proximal Policy Optimization}
\newacronym{ppp}{PPP}{Poisson Point Process}
\newacronym{prach}{PRACH}{Physical Random Access Channel}
\newacronym{prb}{PRB}{Physical Resource Block}
\newacronym{psnr}{PSNR}{Peak Signal to Noise Ratio}
\newacronym{pss}{PSS}{Primary Synchronization Signal}
\newacronym{pucch}{PUCCH}{Physical Uplink Control Channel}
\newacronym{pusch}{PUSCH}{Physical Uplink Shared Channel}
\newacronym{qam}{QAM}{Quadrature Amplitude Modulation}
\newacronym{qci}{QCI}{\gls{qos} Class Identifier}
\newacronym{qoe}{QoE}{Quality of Experience}
\newacronym{qos}{QoS}{Quality of Service}
\newacronym{quic}{QUIC}{Quick UDP Internet Connections}
\newacronym{rach}{RACH}{Random Access Channel}
\newacronym[firstplural=Radio Access Technologies (RATs)]{rat}{RAT}{Radio Access Technology}
\newacronym{rbg}{RBG}{Resource Block Group}
\newacronym{rcn}{RCN}{Research Coordination Network}
\newacronym{rc}{RC}{RAN Control}
\newacronym{rec}{REC}{Radio Edge Cloud}
\newacronym{red}{RED}{Random Early Detection}
\newacronym{renew}{RENEW}{Reconfigurable Eco-system for Next-generation End-to-end Wireless}
\newacronym{rf}{RF}{Radio Frequency}
\newacronym{rfc}{RFC}{Request for Comments}
\newacronym{rfr}{RFR}{Random Forest Regressor}
\newacronym{ric}{RIC}{RAN Intelligent Controller}
\newacronym{rlc}{RLC}{Radio Link Control}
\newacronym{rlf}{RLF}{Radio Link Failure}
\newacronym{rlnc}{RLNC}{Random Linear Network Coding}
\newacronym{rmr}{RMR}{RIC Message Router}
\newacronym{rmse}{RMSE}{Root Mean Squared Error}
\newacronym{rnis}{RNIS}{Radio Network Information Service}
\newacronym{rr}{RR}{Round Robin}
\newacronym{rrc}{RRC}{Radio Resource Control}
\newacronym{rrm}{RRM}{Radio Resource Management}
\newacronym{rru}{RRU}{Remote Radio Unit}
\newacronym{rs}{RS}{Remote Server}
\newacronym{rsrp}{RSRP}{Reference Signal Received Power}
\newacronym{rsrq}{RSRQ}{Reference Signal Received Quality}
\newacronym{rss}{RSS}{Received Signal Strength}
\newacronym{rssi}{RSSI}{Received Signal Strength Indicator}
\newacronym{rtt}{RTT}{Round Trip Time}
\newacronym{ru}{RU}{Radio Unit}
\newacronym{rw}{RW}{Receive Window}
\newacronym{rx}{RX}{Receiver}
\newacronym{s1ap}{S1AP}{S1 Application Protocol}
\newacronym{sa}{SA}{standalone}
\newacronym{sack}{SACK}{Selective Acknowledgment}
\newacronym{sap}{SAP}{Service Access Point}
\newacronym{sc2}{SC2}{Spectrum Collaboration Challenge}
\newacronym{scef}{SCEF}{Service Capability Exposure Function}
\newacronym{sch}{SCH}{Secondary Cell Handover}
\newacronym{scoot}{SCOOT}{Split Cycle Offset Optimization Technique}
\newacronym{sctp}{SCTP}{Stream Control Transmission Protocol}
\newacronym{sdap}{SDAP}{Service Data Adaptation Protocol}
\newacronym{sdk}{SDK}{Software Development Kit}
\newacronym{sdm}{SDM}{Space Division Multiplexing}
\newacronym{sdma}{SDMA}{Spatial Division Multiple Access}
\newacronym{sdn}{SDN}{Software-defined Networking}
\newacronym{sdr}{SDR}{Software-defined Radio}
\newacronym{seba}{SEBA}{SDN-Enabled Broadband Access}
\newacronym{sgsn}{SGSN}{Serving GPRS Support Node}
\newacronym{sgw}{SGW}{Service Gateway}
\newacronym{si}{SI}{Study Item}
\newacronym{sib}{SIB}{Secondary Information Block}
\newacronym{sinr}{SINR}{Signal to Interference plus Noise Ratio}
\newacronym{sip}{SIP}{Session Initiation Protocol}
\newacronym{siso}{SISO}{Single Input, Single Output}
\newacronym{sla}{SLA}{Service Level Agreement}
\newacronym{sm}{SM}{Service Model}
\newacronym{smf}{SMF}{Session Management Function}
\newacronym{smo}{SMO}{Service Management and Orchestration}
\newacronym{sms}{SMS}{Short Message Service}
\newacronym{smsgmsc}{SMS-GMSC}{\gls{sms}-Gateway}
\newacronym{snr}{SNR}{Signal-to-Noise-Ratio}
\newacronym{son}{SON}{Self-Organizing Network}
\newacronym{sptcp}{SPTCP}{Single Path TCP}
\newacronym{srb}{SRB}{Service Radio Bearer}
\newacronym{srn}{SRN}{Standard Radio Node}
\newacronym{srs}{SRS}{Sounding Reference Signal}
\newacronym{ss}{SS}{Synchronization Signal}
\newacronym{sss}{SSS}{Secondary Synchronization Signal}
\newacronym{st}{ST}{Spanning Tree}
\newacronym{svc}{SVC}{Scalable Video Coding}
\newacronym{tb}{TB}{Transport Block}
\newacronym{tcp}{TCP}{Transmission Control Protocol}
\newacronym{tdd}{TDD}{Time Division Duplexing}
\newacronym{tdm}{TDM}{Time Division Multiplexing}
\newacronym{tdma}{TDMA}{Time Division Multiple Access}
\newacronym{tfl}{TfL}{Transport for London}
\newacronym{tfrc}{TFRC}{TCP-Friendly Rate Control}
\newacronym{tft}{TFT}{Traffic Flow Template}
\newacronym{tgen}{TGEN}{Traffic Generator}
\newacronym{tip}{TIP}{Telecom Infra Project}
\newacronym{tm}{TM}{Transparent Mode}
\newacronym{to}{TO}{Telco Operator}
\newacronym{tr}{TR}{Technical Report}
\newacronym{trp}{TRP}{Transmitter Receiver Pair}
\newacronym{ts}{TS}{Technical Specification}
\newacronym{tti}{TTI}{Transmission Time Interval}
\newacronym{ttt}{TTT}{Time-to-Trigger}
\newacronym{tx}{TX}{Transmitter}
\newacronym{uas}{UAS}{Unmanned Aerial System}
\newacronym{uav}{UAV}{Unmanned Aerial Vehicle}
\newacronym{udm}{UDM}{Unified Data Management}
\newacronym{udp}{UDP}{User Datagram Protocol}
\newacronym{udr}{UDR}{Unified Data Repository}
\newacronym{ue}{UE}{User Equipment}
\newacronym{uhd}{UHD}{\gls{usrp} Hardware Driver}
\newacronym{ul}{UL}{Uplink}
\newacronym{um}{UM}{Unacknowledged Mode}
\newacronym{uml}{UML}{Unified Modeling Language}
\newacronym{upa}{UPA}{Uniform Planar Array}
\newacronym{upf}{UPF}{User Plane Function}
\newacronym{urllc}{URLLC}{Ultra Reliable and Low Latency Communications}
\newacronym{usa}{U.S.}{United States}
\newacronym{usim}{USIM}{Universal Subscriber Identity Module}
\newacronym{usrp}{USRP}{Universal Software Radio Peripheral}
\newacronym{utc}{UTC}{Urban Traffic Control}
\newacronym{vim}{VIM}{Virtualization Infrastructure Manager}
\newacronym{vm}{VM}{Virtual Machine}
\newacronym{vnf}{VNF}{Virtual Network Function}
\newacronym{volte}{VoLTE}{Voice over \gls{lte}}
\newacronym{voltha}{VOLTHA}{Virtual OLT HArdware Abstraction}
\newacronym{vr}{VR}{Virtual Reality}
\newacronym{vran}{vRAN}{Virtualized \gls{ran}}
\newacronym{vss}{VSS}{Video Streaming Server}
\newacronym{wbf}{WBF}{Wired Bias Function}
\newacronym{wf}{WF}{Waterfilling}
\newacronym{wg}{WG}{Working Group}
\newacronym{wlan}{WLAN}{Wireless Local Area Network}
\newacronym{osm}{OSM}{Open Source \gls{nfv} Management and Orchestration}
\newacronym{pnf}{PNF}{Physical Network Function}
\newacronym{drl}{DRL}{Deep Reinforcement Learning}
\newacronym{mtc}{MTC}{Machine-type Communications}
\newacronym{mns}{MnS}{Management Services}
\newacronym{ves}{VES}{\gls{vnf} Event Stream}
\newacronym{ei}{EI}{Enrichment Information}
\newacronym{fh}{FH}{Fronthaul}
\newacronym{fft}{FFT}{Fast Fourier Transform}
\newacronym{laa}{LAA}{Licensed-Assisted Access}
\newacronym{plfs}{PLFS}{Physical Layer Frequency Signals}
\newacronym{ptp}{PTP}{Precision Time Protocol}
\newacronym{cbrs}{CBRS}{Citizen Broadband Radio Service}

\newacronym{cif}{CI}{cyberinfrastructure}
\newacronym{sonic}{SONiC}{Software for Open Networking in the Cloud}
\newacronym{ocp}{OCP}{Open Compute Project}
\newacronym{snmp}{SNMP}{Simple Network Management Protocol}
\newacronym{raid}{RAID}{redundant array of independent disks}
\newacronym{nfs}{NFS}{Network File Storage}
\newacronym{ci}{CI}{Continuous Integration}
\newacronym{cd}{CD}{Continuous Deployment}
\newacronym{dtn}{DTN}{Data Transfer Node}

\newacronym{dns}{DNS}{Domain Name Service}
\newacronym{nrpe}{NRPE}{Nagios Remote Plugin Executor}
\newacronym{ldap}{LDAP}{Lightweight Directory Access Protocol}
\newacronym{lan}{LAN}{Local Area Network}
\newacronym{vlan}{VLAN}{Virtual LAN}

\newacronym{ipmi}{IPMI}{Intelligent Platform Management Interface}
\newacronym{tor}{ToR}{Top-of-the-Rack}
\newacronym{lmn}{LMN}{Large Memory Node}
\newacronym{bgp}{BGP}{Border Gateway Protocol}
\newacronym{dhcp}{DHCP}{Dynamic Host Configuration Protocol}
\newacronym{vrf}{VRF}{Virtual Routing and Forwarding}
\newacronym{vpn}{VPN}{Virtual Private Network}
\newacronym{rma}{RMA}{Return Merchandise Authorization}
\newacronym{hpc}{HPC}{High Performance Compute}

\newacronym{nu}{NU}{Northeastern University}
\newacronym{asic}{ASIC}{Application-specific Integrated Circuit}
\newacronym{rdma}{RDMA}{Remote Direct Memory Access}
\newacronym{roce}{RoCE}{RDMA over Converged Ethernet}
\newacronym{ovs}{OVS}{Open vSwitch}
\newacronym{frr}{FRR}{Free Range Routing}
\newacronym{ups}{UPS}{Uninterruptible Power Supply}

\newacronym{ntia}{NTIA}{National Telecommunications and Information Administration}
\newacronym{irb}{IRB}{Institutional Review Board}
\newacronym{doi}{DOI}{Digital Object Identifier}

\newacronym{sdo}{SDO}{Standard-Development Organization}
\newacronym{ose}{OSE}{Open Source Ecosystem}
\newacronym{osc}{OSC}{O-RAN Software Community}
\newacronym{dop}{DOP}{Director of Operations}
\newacronym{pm}{PM}{Program Manager}
\newacronym{excom}{EXCOM}{Executive Committee}
\newacronym{iiot}{IIoT}{Industrial \gls{iot}}
\newacronym{lf}{LF}{Linux Foundation}

\newacronym{wiot}{WIoT}{Institute for the Wireless Internet of Things}

\newacronym{otic}{OTIC}{Open Testing \& Integration Centre}

\newacronym{nofo}{NOFO}{Notice of Funding Opportunity}

\newacronym{onr}{ONR}{Office of Naval Research}
\newacronym{afosr}{AFOSR}{Air Force Office of Scientific Research}
\newacronym{afrl}{AFRL}{Air Force Research Laboratory}
\newacronym{arl}{ARL}{Army Research Laboratory}

\newacronym{arc}{ARC}{Aerial Research Cloud}

\newacronym{mno}{MNO}{Mobile Network Operator}
\newacronym{ct}{CT}{Continuous Testing}
\newacronym{oci}{OCI}{Open Container Initiative}

\newacronym[plural=RANs]{ran}{RAN}{Radio Access Network}
\newacronym{pii}{PII}{Personally Identifiable Information}
\newacronym{cves}{CVEs}{Common Vulnerabilities and Exposures}
\newacronym{cvss}{CVSS}{Common Vulnerability Scoring System}
\newacronym{n-rt-ric}{Near-RT RIC}{Near Real-Time RIC}
\newacronym{non-rt-ric}{Non-RT RIC}{Non Real-Time RIC}
\newacronym{o-cu}{O-CU-CP}{O-RAN Central Unit}
\newacronym{o-cu-cp}{O-CU-CP}{\gls{o-cu} - Control Plane}
\newacronym{o-cu-up}{O-CU-UP}{\gls{o-cu} - User Plane}
\newacronym{o-du}{O-DU}{O-RAN Distributed Unit}
\newacronym{o-ru}{O-RU}{O-RAN Radio Unit}
\newacronym{oran}{O-RAN}{Open Radio Access Network}
\newacronym{sast}{SAST}{Static application security testing}
\newacronym{rbac}{RBAC}{Role-Based Access Control}
\newacronym{cis}{CIS}{Center for Internet Security}
\newacronym{ssrf}{SSRF}{Server-Side Request Forgery}
\newacronym[plural=NFs]{nf}{NF}{Network Function}

\definecolor{critical}{RGB}{163, 28, 18}
\definecolor{high}{RGB}{238, 1, 5}
\definecolor{medium}{RGB}{241, 144, 1}
\definecolor{low}{RGB}{245, 206, 25}
\definecolor{negligible}{RGB}{78, 144, 255}

\makeatletter 
\newcommand{\linebreakand}{%
  \end{@IEEEauthorhalign}
  \hfill\mbox{}\par
  \mbox{}\hfill\begin{@IEEEauthorhalign}
}
\makeatother 

\begin{document}

\title{Securing the Open RAN Infrastructure: Exploring Vulnerabilities in Kubernetes Deployments
\thanks{The authors acknowledge the financial support by the German Federal Ministry of Education and Research -- Bundesministerium für Bildung und Forschung (BMBF), as part of the Project ``6G-RIC: The 6G Research and Innovation Cluster'' (project number 825026). This paper was also partially supported by the U.S. National Telecommunications and Information Administration (NTIA)'s Public Wireless Supply Chain Innovation Fund (PWSCIF) under Award No. 25-60-IF054.}
}

\author{\IEEEauthorblockN{Felix Klement\IEEEauthorrefmark{1}, Alessandro Brighente\IEEEauthorrefmark{2}, Michele Polese\IEEEauthorrefmark{3}, Mauro Conti\IEEEauthorrefmark{2}, Stefan Katzenbeisser\IEEEauthorrefmark{1}}
\IEEEauthorblockN{
\IEEEauthorrefmark{1}Computer Engineering, University of Passau, Passau, Germany\\
\IEEEauthorrefmark{2}Department of Mathematics, University of Padova, Padova, Italy\\
\IEEEauthorrefmark{3}Institute for the Wireless Internet of Things, Northeastern University, Boston, MA, USA\\
Email: \{felix.klement, stefan.katzenbeisser\}@uni-passau.de,\\\{alessandro.brighente, mauro.conti\}@unipd.it, m.polese@northeastern.edu
}}

\IEEEaftertitletext{\vspace{-1\baselineskip}}


\maketitle
\begin{tikzpicture}[remember picture, overlay]
  \node[font=\sffamily\normalsize, yshift=-0.7cm, text centered, text width=\paperwidth, anchor=north west] at (current page.north west) {%
This paper has been accepted for publication at the 10th IEEE International Conference on Network Softwarization (NetSoft2024) 
  };
\end{tikzpicture}


\begin{abstract}
In this paper, we investigate the security implications of virtualized and software-based Open 
\gls{ran} systems, specifically focusing on the architecture proposed by the O-RAN ALLIANCE and O-Cloud deployments based on the \gls{osc} stack and infrastructure. 
Our key findings are based on a thorough security assessment and static scanning of the \gls{osc} Near Real-Time \gls{ric} cluster. We highlight the presence of potential vulnerabilities and misconfigurations in the Kubernetes infrastructure supporting the \gls{ric}, also due to the usage of outdated versions of software packages, and provide an estimation of their criticality using various deployment auditing frameworks (e.g., MITRE ATT\&CK and the NSA CISA). In addition, we propose methodologies to minimize these issues and harden the Open RAN virtualization infrastructure. These encompass the integration of security evaluation methods into the deployment process, implementing deployment hardening measures, and employing policy-based control for RAN components. We emphasize the need to address the problems found in order to improve the overall security of virtualized Open RAN systems.
\end{abstract}


\glsresetall


\section{Introduction}

The Open \gls{ran} paradigm is moving next-generation wireless networks toward systems which are more flexible, programmable, and can be customized and optimized to support new use cases through data-driven intelligent control. Open \gls{ran} solutions, and their implementation as part of the O-RAN ALLIANCE specifications, transition the \gls{ran} toward softwarized solutions, extending the programmability in domains which have usually been associated with custom silicon and dedicated circuits~\cite{polese2023understanding,abdalla2022toward}. 

The adoption of software in the \gls{ran} increases the level of programmability of the full protocol stack, making it easier to interact programmatically with Open RAN base stations. This, in turn, can be leveraged to expose telemetry and performance metrics from the \gls{ran}, collect them at a large scale in controllers at the edge of the network via the \glspl{ric}, and apply data-driven techniques for the intelligent optimization of the stack~\cite{habib2023intent}. At the same time, leveraging software in the \gls{ran} allows for a faster innovation cycle, driven by the possibility of running continuous integration, deployment, and testing processes. Finally, it also increases the diversity of the supply chain as it lowers that barrier to entry in the cellular market. A software-first \gls{ran} requires proper support by the \gls{ran} infrastructure, i.e., a set of virtualization solutions, automation pipelines, and hardware accelerators. This is generally referred to as the \emph{O-Cloud} in the O-RAN ALLIANCE architecture. Virtualization and software-first infrastructure can enable dynamic scaling of compute resources to accommodate and tailor the network deployment to users and traffic requirements, toward a more energy efficient \gls{ran}. It also allows for multi-tenant \gls{ran} data centers, e.g., as in a neutral host environment with multiple operators sharing the same physical infrastructure to reduce costs~\cite{bonati2023neutran}. 
At the same time, a virtualized environment introduces more parameters to configure and tune, additional software components (e.g., multiple software layers between the application and the hardware appliance), and a set of more heterogeneous workloads. This translates into a larger threat surface that can exploited by malicious attackers, either internal or external, or can impact the network performance because of misconfigurations~\cite{wg11ocloud}. 

This paper takes a first step toward understanding threats and quantitatively profiling the vulnerabilities that virtualization introduces in the O-Cloud, focusing specifically on a micro-services-based architecture for the O-RAN \gls{ric} implemented by the \gls{osc}~\cite{i_release}. 
Compared to prior literature on security in Open \gls{ran} systems~\cite{groen2023implementing, liyanage2022open, abdalla2022toward, ramezanpour2022intelligent, mimran2022evaluating}, we focus on an assessment of how the software vulnerabilities in underlying virtualization solutions (specifically, the Kubernetes platform~\cite{near-rt-osc}) impact the services that support the O-RAN \gls{ric}. Our preliminary analyses unveil a substantial quantity of insecure components, configurations, and software dependencies within the \gls{n-rt-ric} cluster presently employed by the \gls{osc}. We leverage static scanning and combine it with a quantitative assessment methodology based on multiple deployment auditing frameworks, including MITRE ATT\&CK, National Security Agency (NSA) and Cybersecurity and Infrastructure Security Agency (CISA) report \cite{cisa}, and the \gls{cis} CIS-v1.23-t1.0.1 \cite{cis}. Our analysis shows 792 vulnerabilities within the \gls{n-rt-ric} and 70 \gls{cves} in the currently used platform versions for the virtualized deployment.


This work testifies to how openness allows for clarifying the attack surface of \gls{ran} systems, a key advantage compared to security-by-obscurity adopted in previous \gls{ran} deployments. We commend the efforts of the \gls{osc} in providing an open-source reference framework for the \gls{ric}, and provide suggestions on how to integrate security assessment methodologies and the general hardening of deployments in the software development lifecycle. We also intend to share the discoveries from this paper with the OSC and address certain issues through the submission of pull requests in the relevant code repositories.
We believe that this analysis can spark further research and focus on securing virtualized and software-based Open RAN systems, a key step toward deploying open, programmable, and intelligent networks that are reliable, resilient, and leverage the best practices of cloud security. 

The remainder of the paper is organized as follows. In Sec.~\ref{sec:ocloud}, we review the O-RAN architecture, O-Cloud, and virtualization solutions. In Sec.~\ref{sec:threat}, we discuss the threat model considered in the paper, combining O-RAN notions and Kubernetes systems. In Sec.~\ref{sec:assessment_methods}, we present security assessment methods, and we present results on their application to the \gls{osc} \gls{ric} software in Sec.~\ref{sec:concerns}. Finally, we discuss best practices in Sec.~\ref{sec:best-practices} and conclude the paper in Sec.~\ref{sec:conclusions}.

\section{O-RAN Architecture and Deployment Methodologies}
\label{sec:ocloud}
In the following section, we briefly introduce the \gls{oran} architecture and explore the various deployment options available.

\begin{figure}[t]
    \centering
    \includegraphics[width=\columnwidth]{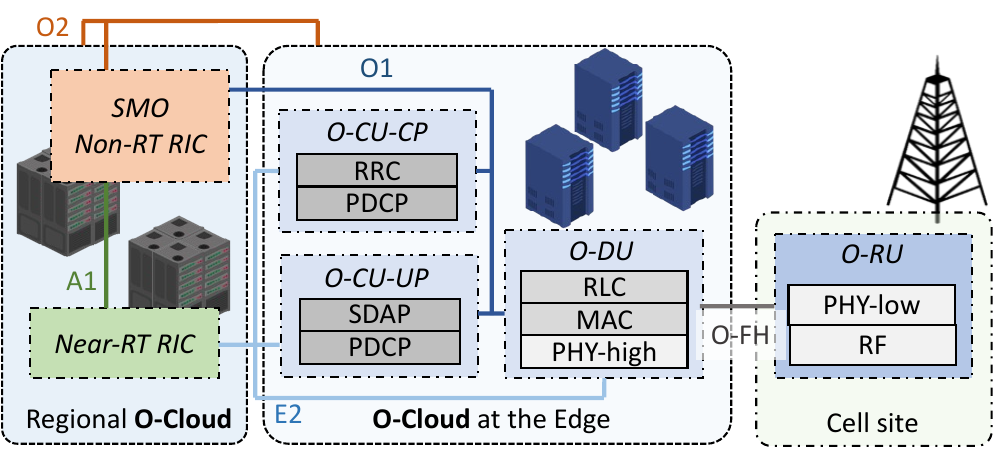}
    \vspace{-2em}
    \caption{Disaggregated O-RAN architecture with open interfaces and a typical deployment across different data centers implementing the O-RAN O-Cloud and a proprietary cell site.}
    \label{fig:architecture}
    \vspace{-1.7em}
\end{figure}

\subsection{O-RAN architecture}
The O-RAN architecture embodies the principles of disaggregation, intelligence, and programmability of the Open \gls{ran} paradigm. This translates into cellular networks which are based on software and are highly automated, and rely on cloud-based or edge computing platforms as essential components of the overall infrastructure. Figure~\ref{fig:architecture} shows a high-level logical diagram of an O-RAN deployment with softwarized components deployed in multiple data centers implementing the O-RAN ALLIANCE O-Cloud, and a proprietary cell site with radio components and front-end, i.e., the \gls{o-ru}~\cite{oranwpusecases}. The O-RAN O-Cloud is a collection of physical infrastructure and software that provides the necessary abstractions and computing power to execute softwarized \gls{ran} workloads. These include the \gls{smo} of the overall network, which also hosts the \gls{non-rt-ric}, the first of two \glspl{ric} that can host custom applications (i.e., rApps) for network management and optimization. The second \gls{ric} is the \gls{n-rt-ric}, which has a direct interface to the \gls{ran} for near-real-time, data-driven resource management and hosts xApps. Finally, the \gls{ran} base stations are disaggregated into the \gls{o-cu}, itself split into a user plane and a control plane function, and the \gls{o-du}. The \gls{smo} orchestrates and manages the deployment of services and solutions in the O-Cloud through the O2 interface, as shown in the top part of Fig.~\ref{fig:architecture}. The specifications on the O-Cloud~\cite{oran-wg6-o-cloud} focus on high-level abstractions rather than mandating a specific technology that needs to be adopted to implement the O-Cloud itself. Nonetheless, the industry has widely adopted microservices implemented through containers as the technical solution to deploy most O-RAN softwarized workloads. In \cite{klement_jsac_oran}, an empirical threat analysis method was used to illustrate that almost none  of the threat classes predefined by the ALLIANCE exist in the O-Cloud without a critical vulnerability. In the upcoming paragraphs, we therefore examine and discuss the contributions of Docker and Kubernetes to enhance RAN deployments and their associated security considerations.


\subsection{Docker as a container, Kubernetes for orchestration}
\vspace{-0.3em}
The complexity of \gls{ran} systems has led to ongoing efforts to transition from challenging-to-manage monolithic approaches to more adaptable, service-oriented solutions based on atomic \glspl{nf}. As explained earlier, the virtualized \gls{ran} has become essential, particularly since the inception of the O-\gls{ran} initiative. Docker containers play a key role in implementing a software-defined mobile network. To ensure the overall security of the containers, their configurations must be secure and consistently kept up to date. Achieving this can often be a substantial task, given that complex systems are constructed from a multitude of such configurations.

In a virtualized \gls{ran} \glspl{nf} are executed in Docker containers within a server cluster using Kubernetes as an orchestrator. The use of Kubernetes facilitates the monitoring and management of \glspl{nf} and thus the dynamic scaling of resources and services to meet changing O-\gls{ran} requirements. As a result, a more flexible and more efficient radio network enables the operator to offer its users a better experience. The results are overall cost savings, improved scalability, and increased agility of the network. Although the advantages of the Kubernetes ecosystem are clear, there are other challenges related to general system security in addition to high performance and latency, e.g., in edge data centers. The use of default configurations can prioritize flexibility over security, which can lead to vulnerabilities \cite{cisa, cis}. Inadequate pod security policies and network vulnerabilities can lead to unauthorized access, service disruption, and data disclosure \cite{k8stop10}. Human error is another important factor to consider, as the complicated nature of Kubernetes increases the likelihood of misconfigurations or oversights. To address these concerns, best practices must be implemented within an O-\gls{ran} deployment, components must be regularly updated and security checks must be performed. Using security tools and applying a defense-in-depth strategy are essential towards improving the overall Kubernetes cluster security.
\vspace{-0.8em}
\section{Threat Model}
\label{sec:threat}
In this section, we present an attacker model and explore potential threat vectors that could be exploited by a malicious entity.

\vspace{-0.3em}
\subsection{Attacker Model} 
\vspace{-0.3em}
We consider an attacker exploiting vulnerabilities in the cloud-based components of \gls{oran}. The attacker can be an insider attacker (i.e., an authorized user of the system) or an external attacker (i.e., a non-legitimate user of the system). The attacker may have different motivations to launch the attack, for instance, gaining access to sensitive data for commercial purposes, disrupting the network due to hacktivism or being a competitor operator, or gaining access to privileged functions they do not own and hence control the network. Insider attackers may also be employees bribed by external actors or enraged with the current employer, hence attacking the system from the inside. 

\vspace{-0.3em}
\subsection{Threat Vectors}
\vspace{-0.3em}
In this paper, we focus on the software component's current security posture of \gls{osc} and refer the reader to the \gls{oran} WG 11 technical specification for details on hardware-related attacks~\cite{wg11ocloud}.
As \gls{osc} leverages Kubernetes for orchestration, we refer to Kubernetes security guidelines, i.e., OWASP Kubernetes Top 10~\cite{k8stop10}, the NSA and CISA Kubernetes hardening guide \cite{cisa}, the MITRE ATT\&CK, and the CIS Kubernetes benchmark \cite{cis}. Due to the lack of space, we can not provide the full details on the possible attacks and their implementations. However, based on the aforementioned guidelines, we summarize the most relevant threat vectors as weak Authentication and Access Control, lack of Network Segmentation and Isolation, Supply Chain vulnerabilities, and use of Outdated Components.

In case of authentication and authorization misconfigurations, the attacker can gain access to restricted resources with capabilities they should not have (e.g., writing permissions), configuration secrets, resource configuration, or impersonate a legitimate user. In \gls{oran}, attackers might gain control over virtual network functions, configurations dealing with network neutrality, confidential data of other operators/infrastructure, and launch resource exhaustion attacks.
Broken network segmentation and isolation imply that the attacker can move inside the Kubernetes deployment. For example, an attacker may access pods belonging to other users. In \gls{oran}, this might lead to, among others, data theft, incomplete termination of network functions, attacks on internal network services, and false resource advertisements.

Pods can be configured to run containers, which should be conveyed from a dedicated repository via a supply chain. Uploading vulnerable or malicious containers represents a severe threat in a Kubernetes deployment. In \gls{oran}, supply chain vulnerabilities might lead to attackers exploiting misconfigurations in the container to gain privileged access to network functions and sensitive user data, or might be an entry point for attackers exploiting lateral movements in the infrastructure. 
Finally, old vulnerable Kubernetes versions might still be available and used. An attacker can hence leverage known vulnerabilities in such configurations. 
\section{Assessment methods}
\label{sec:assessment_methods}

In the following, we discuss our security evaluation approaches within the context of \glspl{ran} and the specific security attributes they assess.

\vspace{-0.3em}
\subsection{Static Scanning}
\label{static_scanning}
\vspace{-0.3em}
\gls{sast} is a methodology employed to analyze static code to uncover potential weaknesses and existing vulnerabilities within the current codebase. It plays a crucial role in code tests associated with the actual business logic. Additionally, it is commonly utilized for validating infrastructure code, such as Docker files. Previous studies \cite{9582243} even show the importance of establishing a connection to runtime security through the use of \gls{sast}. 

\begingroup
\centering
\setlength{\tabcolsep}{7pt} 
\renewcommand{\arraystretch}{1.3} 
\begin{table*}[t]
  \caption{Enumeration of Near-Realtime RIC containers and their counted vulnerabilities and misconfigurations}
  \vspace{-1.5em}
  \label{tab:vulnerabilities_listing}
  \begin{center}
    \begin{tabular}{|l|l|l||l|l|l|l|l||l|l|l|l|l|}
        \multicolumn{3}{c}{Generic Infos} & \multicolumn{5}{c}{Vulnerabilities} & \multicolumn{5}{c}{Misconfigurations} \\ 
    \cmidrule(lr){1-3}\cmidrule(lr){4-8}\cmidrule(lr){9-13}
    \hline
        \multicolumn{1}{|c|}{{Container Name}} & \multicolumn{1}{c|}{Registry} & \multicolumn{1}{c||}{Image Tag}             & {\cellcolor{critical}{C}} & {\cellcolor{high}{H}} & {\cellcolor{medium}{M}} & {\cellcolor{low}{L}} & {\cellcolor{negligible}{N}} & {\cellcolor{critical}{C}} & {\cellcolor{high}{H}} & {\cellcolor{medium}{M}} & {\cellcolor{low}{L}} & {\cellcolor{negligible}{N}} \\ \hline\hline
        ricplt-dbass-redis & nexus3.o-ran-sc.org:10002 & ric-plt-dbaas:0.6.2   & 6        & 14   & 26     & 2   & 0     & 0 & 1  & 3  & 9  &  0         \\ \hline
        influxdb2          & Docker.io                 & influxdb:2.2.0-alpine & 10       & 44   & 28     & 2   & 0     & 0 & 1  & 3  & 9  &  0         \\ \hline
        ricplt-e2term      & nexus3.o-ran-sc.org:10002 & ric-plt-e2:6.0.3      & 0        & 0    & 30     & 31  & 13    & 0 & 1  & 3  & 9  &  0    \\ \hline
        ricplt-rtmgr       & nexus3.o-ran-sc.org:10002 & ric-plt-rtmgr:0.9.4   & 0        & 10   & 119    & 43  & 19    & 0 & 1  & 3  & 9  &  0     \\ \hline
        ricplt-e2mgr       & nexus3.o-ran-sc.org:10002 & ric-plt-e2mgr:6.0.1   & 0        & 4    & 115    & 43  & 19    & 0 & 1  & 3  & 9  &  0         \\ \hline
        ricplt-submgr      & nexus3.o-ran-sc.org:10002 & ric-plt-submgr:0.9.5  & 0        & 10   & 119    & 43  & 19    & 0 & 1  & 3  & 9  &  0     \\ \hline
        ricplt-appmgr      & nexus3.o-ran-sc.org:10002 & ric-plt-appmgr:0.5.7  & 0        & 8    & 36     & 24  & 19    & 0 & 1  & 3  & 9  &  0     \\ \hline
        ricplt-a1mediator  & nexus3.o-ran-sc.org:10002 & ric-plt-a1:3.1.1      & 0        & 9    & 8      & 8   & 7     & 0 & 1  & 3  & 9  &  0     \\ \hline
    \end{tabular}
  \end{center}
   \begin{center}
    Vulnerabilities \& Misconfiguration Scores:  \textcolor{critical}{C $\equiv$ Critical}, \textcolor{high}{H $\equiv$ High}, \textcolor{medium}{M $\equiv$ Medium}, \textcolor{low}{L $\equiv$ Low}, \textcolor{negligible}{N $\equiv$ Negligible}
    \end{center}
  \vspace{-2.8em}
\end{table*}
\endgroup


\subsection{Deployment Auditing}
\vspace{-0.3em}
\label{deployment_auditing}
Regular deployment audits are an important part of securing the cluster against new attacks that may not have existed at the time of initial deployment. 
The audits can be divided into different benchmarks: i)
The benchmark as a single entity, such as that of \gls{cis}, contains a series of globally recognized and consensus-based best practices, and ii) the compliance score as a benchmark.
This complements individual risk scores, which are often an illusory concept and inconsistent between different frameworks. 
The compliance score provides a quantifiable measure of overall security about a set of specific frameworks. The compliance status percentage is calculated by averaging the control compliance scores of all controls within a single framework. 
Frequently used methodologies for mapping this value are, for example, the \gls{cis} as mentioned above, MITRE ATT\&CK, SOC2, DevOpsBest and the NSA CISA \cite{cisa}.

\subsection{Penetration Testing}
\vspace{-0.3em}
With penetration testing, ethical hackers mimic the tactics of malicious actors to uncover potential vulnerabilities that might otherwise go undetected by traditional security measures.
This proactive assessment is important to understand a system's vulnerability to various cyber threats. This type of security testing goes beyond the mere identification of vulnerabilities and provides a comprehensive assessment of the existing security protocols and mechanisms within, for example, an O-RAN deployment. It is important to emphasize that penetration testing is not a one-time event, but an iterative process that evolves as the threat landscape changes. This makes it an important pillar of proactive defense against security threats. 
\vspace{-0.3em}
\subsection{Runtime Security}
\vspace{-0.3em}
The last of the four pillars for the secure operation of O-RAN networks is active runtime security. This is about continuously detecting unexpected behavior, configuration changes, intrusions, and data theft in real-time. Intrusion Detection Systems (IDS) are used in the conventional sense to detect anomalies during runtime. However, these must be complemented through additional tools to validate system-critical configurations and role and rights distributions at iterative intervals, for instance.

\vspace{-0.5em}
\section{Security Concerns}
\label{sec:concerns}
In the forthcoming subsections, we discuss security issues present in the existing open-source implementation of the \gls{osc}. The concerns outlined in Section \ref{outdated_versions} pertain not to a specific component but to the overall implementation status. In Section \ref{scanning_results}, we analyze the \gls{n-rt-ric}, concentrating on a particular virtualized component. This component is deemed crucial from a system-critical perspective and, as such, serves as a potential initial target for attacks. Moreover, other components can be analyzed and evaluated using the same methodologies explained in Section \ref{sec:assessment_methods}. 

It is crucial to note that the current focus of the \gls{osc} is on implementing a functional version of the O-RAN specification. Consequently, the issues we highlight may not be the community's immediate priority, as their main emphasis lies in ensuring the proper functionality of their implementation.

\vspace{-0.3em}
\subsection{Outdated Versions}
\label{outdated_versions}
\vspace{-0.3em}
One of the main concerns is the fact that the official documentation and scripts used to install the dependencies contain very old versions that are often no longer supported. Notably, within the ric-plt-ric-dep repository, the installation script installs Kubernetes 1.16.0 from 2019, currently associated with 23 publicly available \gls{cves}. These vulnerabilities span a \gls{cvss} rating range between 3.0 and 8.8, encompassing potential threats like directory traversal, \gls{ssrf}, Open Redirect, Improper Input Validation, and Denial of Service. Additionally, the referenced Kubernetes Container Network Interface (CNI) version 0.7.5 is susceptible to 9 \gls{cves} with a \gls{cvss} rating range between 7.5 and 8.2, incorporating vulnerabilities like \gls{ssrf}, Infinite Loop, and Resource Exhaustion. The Docker version specified as 20.10.21 is currently exposed to 31 \gls{cves} with \gls{cvss} ratings ranging between 3.3 and 9.8. These issues include concerns such as Improper Certificate Validation, Integer Overflows, and Resource Exhaustion. Lastly, Helm 3.5.4, set to be installed, carries 7 released \gls{cves} with a \gls{cvss} rating range between 4.3 and 8.6, featuring vulnerabilities like Denial of Service, Information Leakage, or Memory Corruption. While acknowledging the likelihood of telecommunications operators resolving and updating such outdated versions in an O-RAN deployment, it remains a security risk for the \gls{osc} to advocate the use of these versions in its documentation and tutorials. Entities unaware of the security risks associated with outdated versions, and consequently neglecting version checks, may unintentionally deploy insecure networks vulnerable to various malicious entry points.

\vspace{-0.5em}
\subsection{Scanning Results}
\label{scanning_results}
\vspace{-0.3em}
For the following assessment, we deploy the \gls{n-rt-ric} in a cluster using the latest version of Kubernetes and employ the methodologies we outline in Section \ref{static_scanning} and \ref{deployment_auditing}. In our analysis, we observe a cumulative total of 792 vulnerabilities, covering a range from critical to low impact ratings, summarized in Table~\ref{tab:vulnerabilities_listing}. The Kubernetes cluster in the "ricplt" namespace demonstrates an average compliance score of 78\% for NSA CISA, 76\% for MITRE ATT\&CK, and 71\% for \gls{cis}-v1.23-t1.0.1. This overall assessment underscores a significant imperative for security measures. Notably, addressing the 16 critical vulnerabilities is paramount, with 10 of them enabling remote code execution. Presently, 13 of these critical vulnerabilities are actionable. 

Certainly, it is important to acknowledge that these vulnerabilities need to be actively exploited by an adversary to pose a threat, and their mere existence does not imply danger. Nevertheless, given that a Kubernetes cluster is a favored target for attacks, it becomes crucial to prevent any known vulnerabilities from being exposed. Additionally, we have identified certain misconfigurations within the cluster, which are issues easily rectified through proper configuration of containers and clusters. The primary challenges encountered include: No resources memory and CPU limits, List Kubernetes secrets, Allow privilege escalation, Anonymous access enabled, Applications credentials in configuration files.
Preventing such issues is a straightforward process, as tools typically offer predefined solutions.


\section{Best Practices}
\vspace{-0.3em}
\label{sec:best-practices}

\subsection{Integration of evaluation methods into the deployment}
\vspace{-0.3em}
Ensuring comprehensive security for the \gls{ran} involves the integration of security assessments throughout all deployment phases. This is crucial for validating security measures and ensuring compliance with specified criteria, such as those for newly uploaded Docker images. Established tools can be utilized to examine configurations, particularly for the potential exposure of sensitive information, including patterns related to \gls{pii}. Additionally, it is fundamental to conduct thorough examinations on the container registry to identify and address potential open \gls{cves}.
Maintaining encrypted transmission for individual artifacts is a key necessity, ensuring basic security across all provisioning phases.
Furthermore, it is crucial to actively incorporate the aspects outlined in Section \ref{sec:assessment_methods} into the deployment process. This integration can be achieved through the implementation of \gls{ci} and \gls{cd}. 
\vspace{-0.7em}

\subsection{Deployment hardening}
\vspace{-0.3em}
This includes securing the Kubernetes API server, through which a malicious actor could cause a lot of damage to the environment. Furthermore, security context hardening is essential, as root users are used by default when pods are started and nothing else is stored in the configuration; this also includes the use of pod security policies. Configuring Kubernetes network policies to control communication between pods and services within the cluster helps to prevent unauthorized access and secure the communication channels.


%
\section{Conclusions}
\label{sec:conclusions}
\vspace{-0.3em}
 
This paper leverages some of the most promising methodologies for assessing security in the virtualized configuration, deployment, and operation of O-Cloud instances within O-RAN deployments. We identified several issues, yet they can be resolved through appropriate and tailored configurations or by updating affected dependencies. A significant challenge in assessing the security status of an O-RAN deployment lies in the absence of a singular deployment model. Currently, numerous initiatives reference and employ varying levels of open-source implementations. Consequently, at this developmental stage, establishing a definitive status for O-RAN, crucial for making relative comparisons, proves impractical. Nevertheless, we remain optimistic about the feasibility of formulating standardized approaches to evaluate these challenges accurately. 

\bibliographystyle{IEEEtran}
\bibliography{biblio}

\end{document}